\documentstyle[aps,twocolumn]{revtex}
\begin{document}
\draft
\title{Vector-like Einstein's equations for $D$-dimensional spherical 
gravity with $(D-2)$-dimensional sphere} 
\author{V.A.Berezin}
\address{Institute for Nuclear Research of the Russian Academy of Sciences,\\
60th October Anniversary Prospect, 7a, 117312, Moscow, Russia
e-mail: berezin@ms2.inr.ac.ru}
\date{\today}
\maketitle
\begin{abstract}
We demonstrate how the Einstein's equations for the $D$-dimensional spherical 
gravity can be written in the covariant vector-like form. These equations 
reveal easily the causal structure of curved spherically symmetric 
manifolds and may appear useful in investigation of different models of brane 
universes.
\end{abstract}

A general spherically symmetric $D$-dimensional space-time is a direct product 
of a two-dimensional pseudo-Euclidean manifold $M_2$ and $(D-2)$-dimensional 
sphere $S_{D-2}$, and its line element can be written in the form 

\begin{equation}
\label{ds2}
ds^2 = g_{\mu\nu} dx^{\mu} dx^{\nu} = g_{ik} dx^i dx^k - 
r^2(x^i) d\sigma^2_{D-2} , 
\end{equation}
where Greek indices run values $(0,1,...D-1)$, Latin indices take two 
values $(0,1)$, and $d\sigma^2_{D-2}$ is the line element of the unit 
$(D-2)$-dimensional sphere, 
\begin{equation}
\label{sigma}
\begin{array}{lc}
d\sigma^2_{D-2} = \sum_{A} \gamma_{AA} d\theta^2_A  , \\
\\
\gamma_{11} = 1, &  0 \leq \theta_1 < 2\pi ,\\
\\
\gamma_{BB} = \sin^2{\theta_1} ... \sin^2{\theta_{B-1}},\\ 
0 \leq \theta_B < \pi , & 2 \leq B \leq D-2 .
\end{array}
\end{equation} 
The radius of the sphere $r(x^i)$ is an invariant function under the 
diffeomorphism on $M_2$.

The three metric coefficients $g_{ik}$ of a two-dimensional manifold can, 
in general, be subject to two coordinate (gauge) conditions. Thus, we are 
left with, essentially, one function defining the metric structure on $M_2$. 
Such a function can be chosen as a invariant (under diffeomorphism on $M_2$). 
This is clear from the well known fact that the general two-dimensional 
metric is (locally) conformally flat, and the conformal factor is an 
invariant. 

But we make another choice. As a second (the radius $r(x^i)$ is the first one) 
invariant we will use a squared normal vector to the surfaces of constant 
radius $r(x^i) = const.$, 
\begin{equation}
\label{delta} 
\Delta = g^{ik} R_{,i} R_{,k} , 
\end{equation}  
where ``$,i$'' denotes a partial derivative with respect to corresponding 
coordinate on $M_2$. It is important that these two invariant functions, 
$r$ and $\Delta$, reveal easily the global geometry of the complete 
two-dimensional manifold, especially its causal structure. In particular, 
the surfaces $\Delta = 0$ define apparent horizons, separating regions in the 
space-time under consideration where $\Delta , 0$ and surfaces of constant 
radius are timelike ($R$-regions, radius can be chosen as a spatial 
coordinate), and those with $\Delta > 0$ ($T$-regions. surfaces $r = const.$ 
are spacelike, and radius can be chosen only as a time 
coordinate) \cite{igor}. 

Let us write down the $D$-dimensional Einstein's equations ($G_D$ - is the 
$D$-dimensional Newtonian constant)
\begin{equation}
\label{G1}
G_{\mu\nu} = R_{\mu\nu} - \frac{1}{2} g_{\mu\nu} R = 8 \pi G_D T_{\mu\nu} , 
\end{equation}
where 
\begin{equation}
\label{R1}
R_{\mu\nu} = \partial_{\lambda} \Gamma^{\lambda}_{\mu\nu} - 
\partial_{\nu} \Gamma^{\lambda}_{\mu\lambda} + 
\Gamma^{\lambda}_{\mu\nu} \Gamma^{\sigma}_{\lambda\sigma} - 
\Gamma^{\lambda}_{\mu\sigma} \Gamma^{\sigma}_{\nu\lambda}
\end{equation} 
is the Ricci tensor, $R = g^{\mu\nu} R_{\mu\nu}$ is the Ricci (curvature) 
scalar, and 
\begin{equation}
\label{Gamma1}
\Gamma^{\lambda}_{\mu\nu} = \frac{1}{2} g^{\mu\nu} (g_{\sigma\mu,\nu} + 
g_{\sigma\nu,\mu} - g_{\mu\nu,\sigma}) 
\end{equation}
are the metric connections.

According to the decomposition of the metric tensor
\begin{equation}
\label{g}
\begin{array}{lc}
g_{\mu\nu} = (g_{ik} - r^2 \gamma_{AA}) , \\
\\
g^{\mu\nu} = (g^{ik} - \frac{1}{r^2} \gamma^{AA}) , \\
\\
g_{il} g^{lk} = \delta^k_i, & \gamma^{AA} \gamma_{AA} = 1 ,
\end{array}
\end{equation}
the metric connections are naturally decomposed into six groups, 
$\Gamma^l_{ik}, \Gamma^l_{Ak}, \Gamma^l_{AB}, \Gamma^A_{ik}, \Gamma^A_{Bk}, 
\Gamma^A_{BC}$. Only for groups are not identically zero, namely,
\begin{equation}
\label{Gamma2}
\begin{array}{l}
\Gamma^l_{ik} = \frac{1}{2} g^{lp} ( g_{pi,k} + g_{pk,i} - g_{ik,p}),\\
\\
\Gamma^l_{AB} = g^{lp} r r_{,p} \gamma_{AA} \delta^A_B ,\\
\\
\Gamma^A_{Bk} = \frac{r{,k}}{r} \delta^A_B ,\\
\\
\Gamma^A_{BC} = \frac{1}{2} \gamma^{AA} ( \gamma_{BB,C} \delta^A_B + 
\gamma_{CC,B} \delta^C_B - \gamma_{BB,A} \delta^B_C) ,
\end{array}
\end{equation}
where $\delta^B_A$ is the usual Kronecker symbol.

The Ricci tensor $R_{\mu\nu}$ is decomposed into three groups, 
$R_{ik}, R_{Ak}, R_{AB}$, and only $R_{ik}$ and $R_{AB}$ are not identically 
zero. They are 
\begin{equation}
\label{R2}
\begin{array}{l}
R_{ik} = {}^{(2)} R_{ik} - (D-2) \frac{r_{|ki}}{r} \\
\\
R_{AB} = {}^{(S)} R_{AB} + (g^{lp} r r_{|lp} + 
(D-3) g^{lp} r_{,l} r_{,p}) \gamma_{AA} \delta^A_B . 
\end{array}
\end{equation}
Here the vertical line denotes a covariant derivative with respect to 
the metric $g_{ik}$ on $M_2$, ${}^{(2)}R_{ik}$ is the Ricci tensor of $M_2$, 
and ${}^{(S)}R_{AB}$ is the Ricci tensor of the $(D-2)$-dimensional unit 
sphere. The curvature scalar is 
\begin{equation}
\label{RS}
\begin{array}{l}
R = {}^{(2)}R - \\
\\
\frac{1}{r} {}^{(S)}R - \frac{2 (D-2)}{r} g^{ik} r_{|ik} - 
\frac{(D-2)(D-3)}{r^2} g^{ik} r_{,i} r_{,k} ,
\end{array}
\end{equation}
where ${}^{(2)}R$ and ${}^{(S)}R$ are the curvatures of $M_2$ and the unit 
sphere, respectively. For the general $n$-dimensional sphere the mixed 
component Ricci tensor and the curvature scalar are 
\begin{equation}
\label{S1}
\begin{array}{c}
{}^{(n)}R^A_B = (n-1) \delta^A_B ,\\
\\
{}^{(n)}R = n (n-1) . 
\end{array}
\end{equation}
In our case $n = D-2$ and we have, finally 
\begin{equation}
\label{Rf}
\begin{array}{l}
R_{ik} = {}^{(2)}R_{ik} - \frac{(D-2)}{r} r_{|ik} ,\\
\\
\\
R^A_B = \frac{1}{r^2} ((D-3) + g^{lp} r r_{|lp} + 
(D-3) g^{lp} r_{,l} r_{,p}) \delta^A_B ,\\ 
\\
\\
R = {}^{(2)}R - \\
\\
\frac{2(D-2)}{r} g_{ik} r_{|ik} - 
\frac{(D-2)(D-3)}{r^2} g^{ik} r_{,i} r_{,k} - 
\frac{(D-2)(D-3)}{r^2} .
\end{array}
\end{equation}

The Einstein's equations are also decomposed in the the two-dimensional 
$M_2$-part and the angular part. Let us first write the latter one, 
\begin{equation}
\label{G2}
\begin{array}{l}
G^A_B = R^A_B - \frac{1}{2} \delta^A_B R = \\
\\
\frac{(D-3)}{r} g^{ik} r{|ik} + 
\frac{(D-3)(D-4)}{2 r^2} g^{ik} r_{,i} r_{,k}\\
\\ 
+ \frac{(D-3)(D-4)}{2 r^2} - \frac{1}{2} {}^{(2)}R) \delta^A_B = 
 8 \pi G_D T^A_B \delta^A_B .
\end{array}
\end{equation}
Note that all the angular components of the energy-momentum tensor $T^A_A$ 
are equal and invariant under coordinate transformations on $M_2$. 

The $M_2$-components of the Einstein's equations take now the following form
\begin{equation}
\label{G3}
\begin{array}{l}
G_{ik} = R_{ik} - \frac{1}{2} g_{ik} R = - \frac{(D-2)}{r} r_{|ik} + \\ 
\\
\frac{1}{2} g_{ik} (D-2) (\frac{2}{r} g^{lp} r_{|lp} + 
\frac{D-3}{r^2} g^{lp} r_{,l} r_{,p} + \frac{D-3}{r^2}) = \\
\\
8 \pi G_D T_{ik} .
\end{array}
\end{equation}
It is a well known fact that the two-dimensional Einstein's tensor 
${}^{(2)} R_{ik} - 1/2 g_{ik} {}^{(2)} R$ is zero everywhere except, 
maybe, some singular points, so, it disappeared safely from Eqn.(\ref{G3}) 
(but the curvature scalar ${}^2 R$ enters our Eqn.(\ref{G2}) !). 

In order to reach our goal and write the Einstein's equations for the 
spherical $D$-dimensional gravity in a vector-like form let us do the 
following. First we multiply Eqn.(\ref{G3}) for $G^k_i$ by 
$r^2 r_{,k}$ and sum over $k$. We get 
\begin{equation}
\label{V1}
\begin{array}{c}
G^k_i r_{,k} r^2 = (D-2) g^{lp} r_{|lp} r_{,i} r - \\
\\
(D-2) g^{kl} r_{|il} r_{,k} + \frac{(D-2)(D-3)}{2} r_{,i} \Delta + \\
\\
\frac{(D-2)(D-3)}{2} r_{,i} = 8 \pi G_D T^k_i r_{,k} r^2 .
\end{array}
\end{equation}
Here $\Delta$ is our second invariant function introduced earlier. Then, 
we take trace of the Einstein's tensor $Tr (G_{ik}) = G^k_k$ and 
multiply it by $r^2 r_{,i}$ : 
\begin {equation}
\label{V2}
\begin{array}{l}
G^k_k r^2 r_{,i} = \\
\\
(D-2) g^{kl} r_{|kl} r_{,i} r + (D-2)(D-3) r_{,i} \Delta + 
(D-2)(D-3) r_{,i} = \\
\\
8 \pi G_D T r^2 r_{,i} , 
\end{array}
\end{equation}
where $ T = Tr (T_{ik}) = T^k_k$ .  But, 
\begin{equation}
\label{V3}
g^{kl} r_{|il} r_{,k} = \frac{1}{2} \Delta_{,i} .
\end{equation}
Subtracting now Eqn.(\ref{V1}) from Eqn.(\ref{V2}) , we get 
\begin{equation}
\label{V}
\begin{array}{c}
r \Delta_{,i} + (D-3) r_{,i} \Delta + (D-3) r_{,i} = \\
\\
\frac{16 \pi G_D}{D-2} r^2 (T r_{,i} - T^k_i r_{,k}) . 
\end{array}
\end{equation}
This equation can be written in a more elegant form 
\begin{equation}
\label{Vf}
(r^{(D-3)} ( 1+ \Delta))_{,} = 
\frac{16 \pi G_D}{D-2} r^{D-2} (T r_{,i} - T^k_i r_{,k}) . 
\end{equation}
The Eqn.(\ref{Vf}) is just what we called the vector-like form of the 
Einstein's equations for the spherical gravity. 

Till now we obtained two equations. But, there are, in total, four 
Einstein's equations for the $D$-dimensional spherical gravity: 
three of them with $T_{ik}$ in the right hand side, and the fourth one 
contains $T^A_A$. Our Eqn.(\ref{Vf}) is a system of two partial derivative 
equations, their left hand sides are just the partial derivatives of 
some scalar function. Thus, their right hand sides should obey a usual 
integrability condition. It can be shown that such an integrability 
condition together with two equations contained in Eqn.(\ref{Vf}) are 
equivalent to the three of Einstein's equations with $T_{ik}$ in their 
right hand sides. The fourth equation, namely, Eqn.(\ref{G2}), is not
very convenient for the use because of the presence of the 
two-dimensional curvature scalar ${}^2 R$. Instead, we can use the Bianci 
identities that in our case are reduced to 
\begin{equation}
\label{B}
\begin{array}{c} 
(r^{D-2} T^k_i )_{|k} = (r^{D-2})_{,i} T^A_A \\
\\
T^B_{A|B} = 0 \Longrightarrow T^A_{A,B} = 0 .
\end{array}
\end{equation}
Of course, in practice, we need only one (any one) of the two equations 
in the first line. The other one can be rewritten, after eliminating 
$T^A_A$ and making use of the Eqn.(\ref{Vf}) and the abovementioned 
integrability condition, as follows. 
\begin{equation}
\label{end}
r_{|il} T^l_k - r_{|kl} T^l_i = 0 .
\end{equation}
This equation may also appear useful because it does not contain 
the derivatives of the energy-momentum tensor $T^k_i$. 

To my knowledge, the vector-like equations, Eqn.(\ref{Vf}),  
for the spherical gravity in four dimensions were first 
derived in \cite{Frolov}, and in the covariant form the Eqns.(\ref{Vf}) and 
(\ref{B}) were considered in \cite{BKT}.       
\vskip5mm
The author thanks the Russian Foundation for Basic Research for the 
financial support (grant 99-02-18524-a).


\begin{thebibliography}{99}

\bibitem{igor} I.D.Novikov, Commun. Sternberg Astr. Inst., {\bf 132}, 3 
(1964), {\bf 132}, 43 (1964) 
\bibitem{Frolov} M.A.Markov, V.P.Frolov, Teoreticheskaya i matematicheskaya 
fizika, {\bf 3}, 3 (1970) (in Russian) 
\bibitem{BKT} V.A.Berezin, V.A.Kuzmin, V.A.Tkachev, Phys.Rev., 
{\bf D 36}, 2919 (1987)
\end{thebibliography}
\end{document}